\documentclass[11pt,twoside]{article}
\usepackage{asp2014}

\newcommand{\pycft}{PyCF3}

\aspSuppressVolSlug
\resetcounters

\begin{document}

\title{Being nice to the server: Wrapping a REST API for a cosmological distance/velocity calculator with Python}

% full name: Martin Beroiz
\author{Juan~Cabral$^{123}$, Ehsan~Kourkchi$^4$, Martin~Beroiz$^5$, Erik~Peterson$^6$ and Bruno~S\'anchez$^6$}

\affil{$^1$Instituto De Astronom\'ia Te\'orica y Experimental (IATE -- CONICET), C\'ordoba, C\'ordoba, Argentina; \email{jbcabral@unc.edu.ar}}
\affil{$^2$Centro Internacional Franco Argentino de Ciencias de la Informaci\'on y de Sistemas (CIFASIS, CONICET--UNR), Rosario, Santa f\'e, Argentina}
\affil{$^3$Comisi\'on Nacional de Actividades Espaciales (CONAE), Falda del Ca\~nete, C\'ordoba, Argentina}
\affil{$^4$Institute for Astronomy, University of Hawaii, Honolulu, Hawaii, USA}
\affil{$^5$California Institute of Technology, Pasadena, California, USA}
\affil{$^6$Department of Physics, Duke University, Durham, North Carolina, USA}

% \affil{$^2$Institution Name, Institution City, State/Province, Country}
% remove/add as you need

% remove/add authors as you need
\paperauthor{Juan~Cabral}{jbcabral@unc.edu.ar}{0000-0002-7351-0680}{Instituto De Astronom\'ia Te\'orica y Experimental (IATE -- CONICET)}{}{C\'ordoba}{C\'ordoba}{5000}{Argentina}
\paperauthor{Ehsan~Kourkchi}{ekourkchi@gmail.com}{0000-0002-5514-3354}{University of Hawaii}{Institute for Astronomy}{Honolulu}{Hawaii}{96822}{USA}
\paperauthor{Martin~Beroiz}{mberoiz@caltech.edu}{0000-0001-6486-9897}{California Institute of Technology}{LIGO Caltech}{Pasadena}{California}{91125}{USA}
\paperauthor{Erik~Peterson}{erik.r.peterson@duke.edu}{0000-0001-8596-4746}{Duke University}{Department of Physics}{Durham}{North Carolina}{27708}{USA}
\paperauthor{Bruno~S\'anchez}{bruno.sanchez@duke.edu}{0000-0002-8687-0669}{Duke University}{Department of Physics}{Durham}{North Carolina}{27708}{USA}
% remove/add as you need

% leave these next few aindex lines commented for the editors to enable them. Use Aindex.py to generate them for yourself.
% first presenting author should be the first entry for bold-facing the author index page-reference
%\aindex{Beroiz,~M.}
%\aindex{Author2,~S.}
% remove/add as you need

% leave the ssindex lines commented for the editors to enable them, use Index.py to suggest yours
%\ssindex{FOOBAR!conference!ADASS 2020}
%\ssindex{FOOBAR!organisations!ASP}

% leave the ooindex lines commented for the editors to enable them, use ascl.py to suggest yours
%\ooindex{FOOBAR, ascl:1101.010}
  
\begin{abstract}

In this paper we present PyCF3, a python client for the cosmological distance-velocity calculator CosmicFlow-3. The project has a cache and retry system designed with the objective of reducing the stress on the server and mitigating the waiting times of the users in the calculations. 
We also address Quality Assurance code standards and availability of the code.
  
\end{abstract}

\section{Introduction}
\label{section:intro}
Releasing astronomical data means one must provide the community with meaningful selection, transformation, and presentation of query results through adequate tools.
Peculiar galaxy velocity data is no exception, and to make cosmic distance-velocity data truly open and available to the community, web query applications as well as programmatic APIs are needed.

Python has become in recent times the de-facto programming language for data analysis and creation of computer tools in astronomy \citep{greenfield2011python}.
Although Python was conceived as a general purpose language and designed to serve as the ``glue'' between multiple tools and languages \citep{sanner1999python}, it is currently the first choice for scientific programming. Examples of these are Scikit-Learn for machine learning \citep{scikit-learn}, Astropy for astronomy \citep{astropy:2018}, Tensoflow for deep-learning \citep{tensorflow2015-whitepaper}, among others.

In this context, a wide range of utility packages and applications for diverse platforms are publicly available to the community.
A particular case are tools whose functionalities are accessible through some type of web service.
These ``webapps'' are usually employed through a web interface in an HTML page and/or, less frequently in some type of programming interface.

In the latter case, the ``Cosmicflows-3 Distance–Velocity Calculator - CF3'' tools \citep{2020AJ....159...67K} provide access to the data through an API-JSON \citep{tang2021common} which enables the integration of this tool into the Python scientific ecosystem.

In this work we present the complete integration of CF3 into the python scientific stack, and we describe the technologies and the engineering processes involved in the task of guaranteeing correct behavior and minimal stress load in the main CF3 server.

\section{Cosmicflows-3 Distance–Velocity Calculator - CF3}

The Cosmicflows-3 (CF3) package is a collection of two applications, used to calculate the relationships between distances and velocities of galaxies, based on a smoothed version of the velocity field in the local universe \citep{2020AJ....159...67K}:
The first application is the \textbf{NAM: D-V calculator}. This application computes the expectation of distances or velocities based on a smoothed velocity field from the Numerical Action Methods model (NAM) \citep{2017ApJ...850..207S} appropriate for distances shorter than 38~Mpc. The second application, the \textbf{CF3: D-V calculator} computes the expectation of distances or velocities based on smoothed velocity field from the Wiener filter model \citep{2019MNRAS.488.5438G}. This second model is applicable to distances up to 200~Mpc.

As mentioned in Section~\ref{section:intro}, CF3 is a publicly available web application on the internet, both in the form of a web page and a REST API (\url{http://edd.ifa.hawaii.edu/CF3calculator/}). They expose the calculator data in JSON format \citep{ecmajson} through an HTTP protocol. Having this programming interface opens up many possibilities for integrating the tool with various data analysis ecosystems, but it is important to note that properly managing access to remote servers is not an easy task. As an example, sending requests through the internet to perform the same task multiple times can cause unnecessary waits. In the case of CF3, these waits are in the order of seconds per request.

\section{pycf3 - Cosmicflows Galaxy Distance-Velocity Calculator client for Python}

For CF3, we developed a library named \pycft{}, which integrates the web service of both calculators in a simple way. 
\pycft{} also transparently orchestrates a cache to speed up repeated computations, and retries to attenuate network failures.

The Python \pycft{} client exposes two classes, \texttt{pycf3.NAM} and \texttt{pycf3.CF3}. These provide access to each of the project's calculators. Both classes have configurable options, such as which cache to use and the number of retries in case of network failure. The functionality is condensed in two methods to calculate distances or velocities with respect to three coordinate systems.
These methods returns an object of type \texttt{pycf3.Result} which contains all the requested calculation information. As an example, to estimate the speed at a distance of 180~Mpc using an equatorial coordinate system with the \textit{CF3} calculator, the code would be:

\begin{verbatim}
>>> import pycf3
>>> cf3 = pycf3.CF3()
>>> cf3.calculate_velocity(distance=180., ra=187, dec=13)
Result - CF3(distance=180, ra=187, dec=13)
+----------+-----------------+--------------------+
| Observed | Distance (Mpc)  | [180.]             |
|          | Velocity (Km/s) | 12515.699706446017 |
+----------+-----------------+--------------------+
| Adjusted | Distance (Mpc)  | [180.]             |
|          | Velocity (Km/s) | 12940.58481990226  |
+----------+-----------------+--------------------+
\end{verbatim}

For more examples, and the complete list of attributes and the use of the  \texttt{pycf3.Result} object, please check the \pycft{} documentation (\url{https://pycf3.readthedocs.io/}).

\subsection{Cache and retry subsystems}

The direct calls to the CosmicFlow-API are managed by two main subsystems of \pycft, the re-try system and the cache system.
In the case of an unfulfilled request, due to a connection problem, the retry system will rely on the ``requests'' library \citep{chandra2015python} to repeat the call.
The retry system works together with the cache system, whose role is more important since it prevents sending repeated requests for a previously called calculation.
The cache subsystem is implemented on top of the ``disckcache'' (\url{https://pypi.org/project/diskcache/}) package.

\begin{figure}
\begin{center}
\includegraphics[width=0.8\columnwidth]{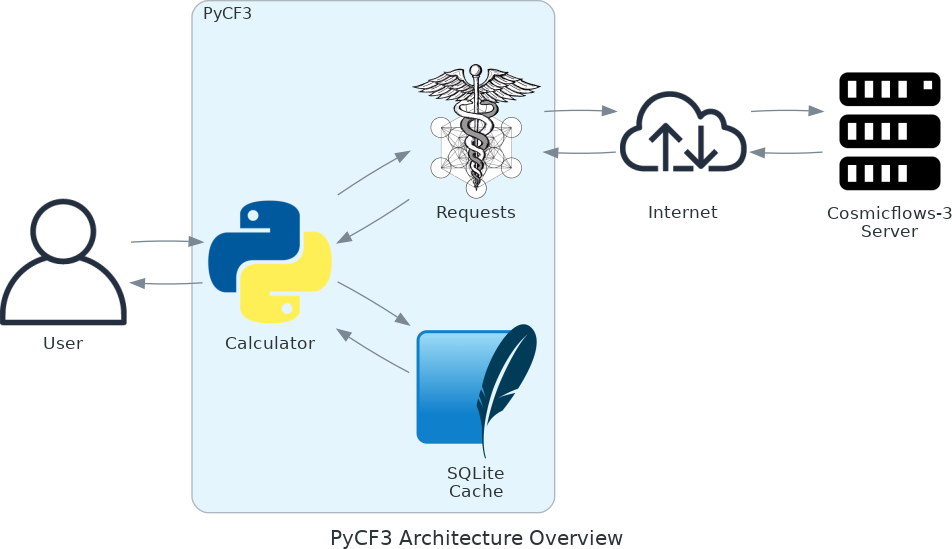}
\caption{
Main parts and libraries that make up the complete \pycft~ calculation pipeline. The main subsystems of the package are enclosed in the blue rectangle, these are the calculator, the cache, and the requests retry system.}
\label{fig:arch}
\end{center}
\end{figure}

The entire life cycle of a \pycft{} request can be summarized as follows:

\begin{enumerate}
\item The user requests the computation of a distance or velocity.
\item Check if the requested computation corresponds to one already stored in the local cache.
  \begin{enumerate}
      \item If the cache does not exist, the computation request is sent to the server and the result is stored in the cache. If the request fails, use the retry subsystem (by default it retries 3 times with waiting times of 300 ms per request).
  \end{enumerate}
\item The result is extracted from the cache and the response is returned to the user.
\end{enumerate}
With this strategy, in addition to reducing the load on the CF3 server, the repeated computation decreases, to only $\sim 3$ or $ \sim 2 ms $. A global view of the \pycft{} architecture can be seen in Figure~\ref{fig:arch}

\section{Quality}

To guarantee the correct functioning of \pycft{}, the project has
150 unit tests, 19 integration tests, validation of styles with flake8 (\url{https://pypi.org/project/flake8/}) and pydocstyle (\url{https://pypi.org/project/pydocstyle/}), and a 97\% code coverage.

All the code is freely available on GitHub (\url{https://github.com/quatrope/pycf3}), along with extensive documentation. The proyect is pip-installable \texttt{pip install pycf3} from PyPI (\url{https://pypi.org/project/pycf3/}).
A continuous integration system (\url{https://travis-ci.com/quatrope/pycf3}) guarantees that each \texttt{git commit} has a correct execution.

\section{Conclusions}
We presented \pycft{}: a tool that enables the access to the Cosmicflows Galaxy Distance-Velocity Calculator CF3 through a Python client, which is able to orchestrate cached results and repeated attempts of querying the server, in order to correctly handle network failures.

In the future we intend to extend the project to include Cosmicflows-4 \citep{kourkchi2020cosmicflows} results as a new calculator, implement units and provide some kind of graphing utilities to facilitate interactive use of the data.

\bibliography{X0-017}

% if we have space left, we might add a conference photograph here. Leave commented for now.
% \bookpartphoto[width=1.0\textwidth]{foobar.eps}{FooBar Photo (Photo: Any Photographer)}

\end{document}